\documentclass[aps,pra,twocolumn,groupedaddress]{revtex4-1}

\usepackage{tabularx,amsmath,amssymb,color}
\usepackage{amsmath}
\usepackage{graphicx}
\usepackage{setspace}
\usepackage{booktabs}
\usepackage{float}
\usepackage{subfigure}

\usepackage{changes}  
\usepackage{lipsum}
\definechangesauthor[name={Benature}, color=orange]{ben}
\definechangesauthor[name={MQL}, color=orange]{mql}
\definechangesauthor[name={Joe}, color=orange]{joe}
\graphicspath{{fig/}}

\usepackage[bookmarks=true,
            colorlinks,
            linkcolor=cyan,
            anchorcolor=cyan,
            citecolor=cyan]{hyperref} 

\def\mm{{\,\mathrm{mm}}}
\def\um{{\,\mu\mathrm{m}}}
\def\V{{\mathrm{V}}}
\def\d{{\mathrm{d}}}

\def\kHz{{\,\text{kHz}}}
\def\Yb{{\text{Yb}}}
\def\LG{{\text{LG}}}

\begin{document}

\title{Design of a novel monolithic parabolic-mirror ion-trap to precisely align the RF null point with the optical focus}

\author{Zhao Wang}
\email{joeshardow@gmail.com}
\author{Ben-Ran Wang}
\author{Qing-Lin Ma}
\author{Jia-Yu Guo}
\author{Ming-Shen Li}
\author{Yu Wang}
\author{Xin-Xin Rao}
\author{Zhi-Qi Huang}
\email{huangzhq25@mail.sysu.edu.cn}
\author{Le Luo}
\email{luole5@mail.sysu.edu.cn}
\affiliation{School of Physics and Astronomy, Sun Yat-Sen University, Zhuhai, Guangdong, China 519082}


\date{\today}

\begin{abstract}

We propose a novel ion trap design with the high collection efficiency parabolic-mirror integrated with the ion trap electrodes. This design has three radio frequency (RF) electrodes and eight direct current(DC) compensation electrodes. By carefully adjusting three RF voltages, the parabolic mirror focus can be made precisely coincident with the RF null point. Thus, the aberration and the ion micromotion can be minimized at the same time. This monolithic design can significantly improve the ion-ion entanglement generation speed by extending the photon collecting solid angle beyond $90\%\cdot4\pi$. Further analysis of the trapping setup shows that the RF voltage variation method relexes machining accuracy to a broad range. This design is expected to be a robust scheme for trapping ion to speed entanglement network node. 
\end{abstract}


\maketitle


\section{Introduction}

Trapped ion quantum computing have been developing rapidly in recent years. Several universal quantum computer prototypes with tens of qubits were demonstrated recently~\cite{wright2019benchmarking,debnath2016demonstration,arute2019quantum,wang2019boson}. In all respects, quantum state preparation, gate operations, measurements, the fidelity and coherence requirements have surpassed the fault-tolerant computing threshold simultaneously in one single quantum system in the same experiment~\cite{PhysRevLett.117.060504, PhysRevLett.117.060505, barends2014superconducting}. In order to employ the quantum advantage in practical applications, further increasing the number of entangled qubits to thousands or millions is one of the next key development steps. 

One of the prominent scalable quantum system methods to use is the distributed quantum network. This idea breaks a large entanglement state with many qubits into several small entanglement states, using photons and interference effects to perform entanglement connections and gate operations~\cite{Duan:2004:STI:2011617.2011618,duan2006probabilistic,duan2010colloquium,Moehring:07,duan2001long}. Each small entangled state will be composed of tens of ions, called a quantum network node, with the function of the local universal quantum computer and the ion-photon interface~\cite{wright2019benchmarking,debnath2016demonstration,stephenson2019high,hucul2015modular}. The final number of entangled qubits is determined by the ratio of qubits coherence time and the successful established photon entangling gate time. The key element of the solution is to decrease the success photon gate time or to improve photon gate success probability. In the two-photon entanglement scheme, the total succeeded ion-ion entanglement probability is proportional to the product of several experimental parameters in each node, i.e. collected solid angle, objective lens transmission(reflection) efficiency, fiber coupling efficiency, detector efficiency,  etc~\cite{Duan:2004:STI:2011617.2011618,duan2006probabilistic,hucul2015modular,stephenson2019high}. Among all factors, the efficiency of photon collection in state-of-the-art ion-ion entanglement experiments is still less than $10 \%$ \cite{hucul2015modular,stephenson2019high}, and is the most restrictive factor to the total entanglement success probability, compared to other experimental parameters. Another key restrictive factor is the single mode fiber(SMF) coupling efficiency. Because a high successful probability of the two-photon interference requires a high degree of the spatial transverse mode overlap. It means that the ion dipole radiation needs to be converted to the Gaussian mode shape of the fiber~\cite{kim2011efficient,sondermann2007design}.

Several previous studies have attacked the collection problem of ion trap systems, such as Fresnel lens~\cite{streed2011imaging}, spherical mirror~\cite{shu2011efficient,noek2010multiscale,hetet2010qed}, parabolic mirror etc.~\cite{chou2017note,maiwald2012collecting,lindlein2007new,luo2009protocols}, cavity \cite{sterk2012photon,casabone2015enhanced,steiner2013single,ballance2017cavity,takahashi2020strong}. The most impressive result is the parabolic mirror method, in which the collection efficiency can be improved to $54\%$\cite{maiwald2012collecting,alber2017focusing}. If using this kind of parabolic mirror trap, the total ion-ion entanglement success probability expected to be improved 29 times higher than trap with a $10 \%$ photon collection efficiency~\cite{debnath2016demonstration,stephenson2019high}. Besides, the reflected photons need to be kept in a pure single transverse mode. This restriction requires the critical spatial coincidence between the parabolic-mirror focus and the RF saddle point~\cite{chou2017note}. The ion at the focal point keeps the collected photon in aberration-free condition, i.e., the residual optical path difference(OPD) should be below Rayleigh criteria\cite{barakat1965rayleigh}. The ion at the RF saddle point is free from micromotion, and behaves like a perfect point light source.

In this work, we propose a novel trap structure, which includes three RF electrodes and eight symmetric DC electrodes. The RF null point can be shifted to be precisely coincident  with the focal point by changing three RF trap electrode voltages. The micromotion can be compensated by carefully adjusting DC voltages. By using metal parabolic mirror surface as the trap electrodes at the same time~\cite{luo2009protocols}, this parabolic mirror trap design can be further integrated and be more compact. The trap structure, which uses purely electric method to align the RF null point with the focus and does not involve any mechanical movable components, is simple and easy to assembly. The coincidence accuracy is determined by the precision and the slew-rate of the RF source. The RF null point can be locked exactly at the focus with real-time feedback if using a high slew-rate and high precision RF voltage source. Further analysis shows that this design has a relaxed requirements of the machining accuracy, thus it potentially constitutes a practical entanglement network node.

This article is organized as follows. Section~\ref{sec:design} describes the geometric structure of the parabolic-mirror trap and the fundamental trap parameters. 
Section~\ref{sec:prop} analyses the optical properties and the trapping laser considerations. Section~\ref{sec:overlap}  demonstrates the key adjustment ability for precise spatial coincidence of the focus and the RF saddle point. In the last section we summarize and conclude.

\begin{figure}[b]
	\subfigure[front view]{
		\includegraphics[width=.48\textwidth]{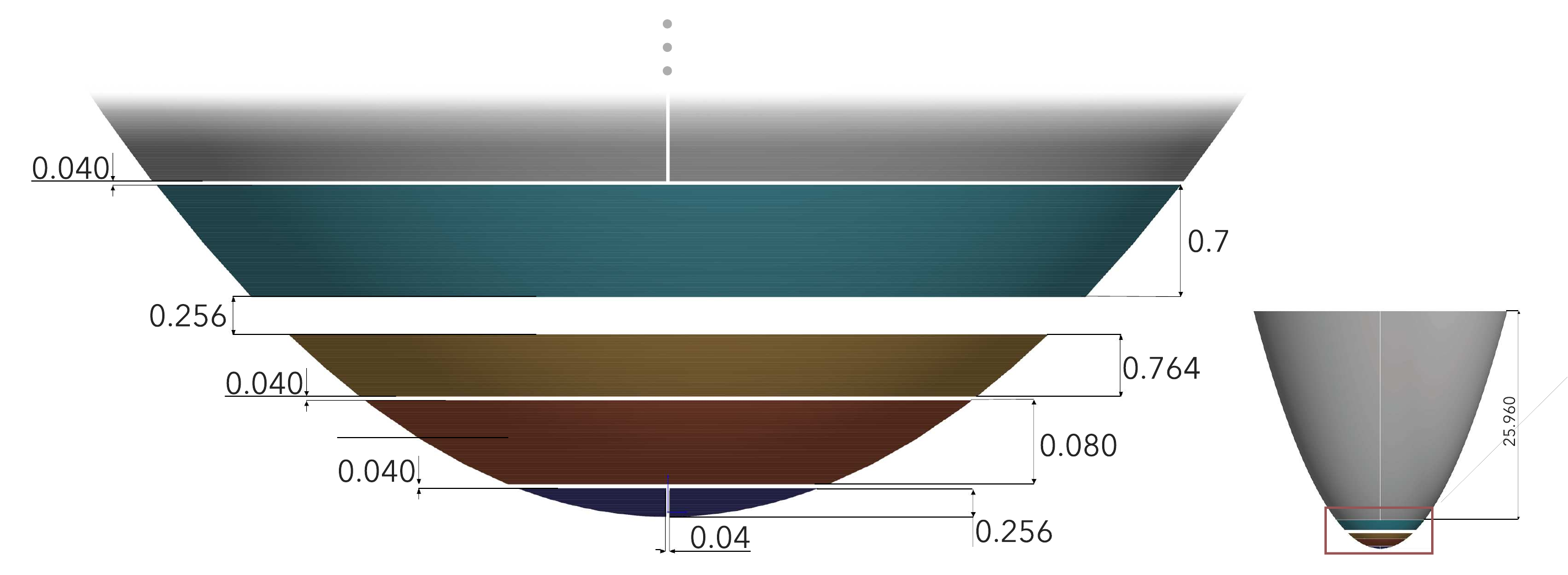}
		\label{frontview}
	}
	\subfigure[top view]{
		\includegraphics[width=.48\textwidth]{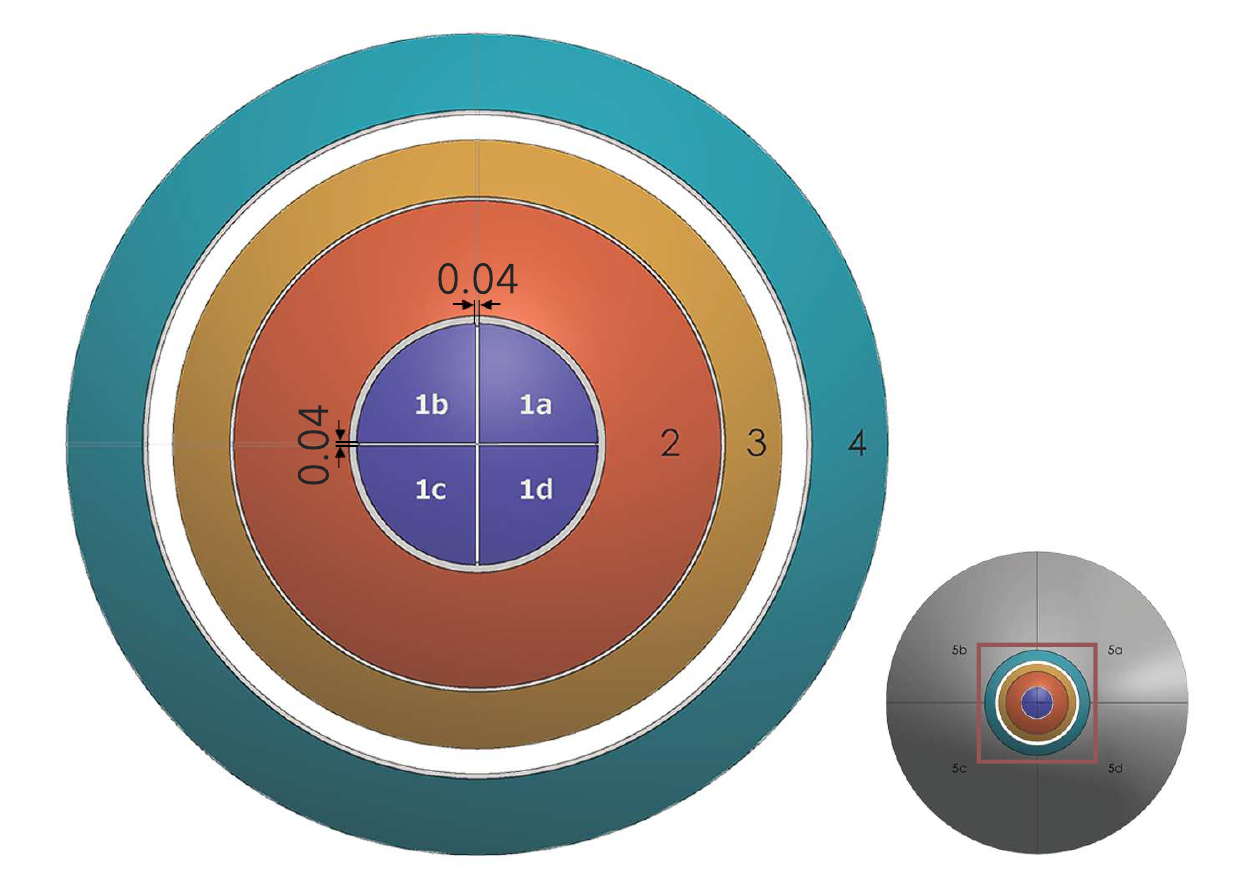}
		\label{topview}
	}%
	\caption{(a)The front view of the monolithic parabolic-mirror trap structure geometry. The geometric dimension is annotated with the original coordinate point of paraboloid apex. The designed paraboloid has a focal length of $f=2.1\,\mathrm{mm}$, with a front opening of $31.5\,\mathrm{mm}$ in diameter, resulting in a depth of  $29.5\,\mathrm{mm}$. The surface of the parabolic mirror is designed as high reflecting aluminum coating and it is divided into 5 separated electrode segments, labeled by numbers 1 to 5. The bottom and top segments(the label 1 segment and the label 5 segment) are DC trap electrodes, the middle three segments(labeled 2-4) are RF trap electrodes. The focus of the parabolic mirror is set in the middle of the gap between segment 3 and segment 4 and it coincides with the RF saddle point. (b)The bottom view. The DC electrodes each(the segment label 1 in a purple color and labeled 5 in a grey color) are further divided into 4 pieces. Each piece is labeled by alphabets, 1a-1d, 5a-5d. The gap dimension is annotated in the figure, along the $x$ and $y$ axes with the original point of the center, paraboloid apex.}
\end{figure}

\section{Trap Design~\label{sec:design}} 

 \begin{figure}[b]
 	\includegraphics[width=8cm]{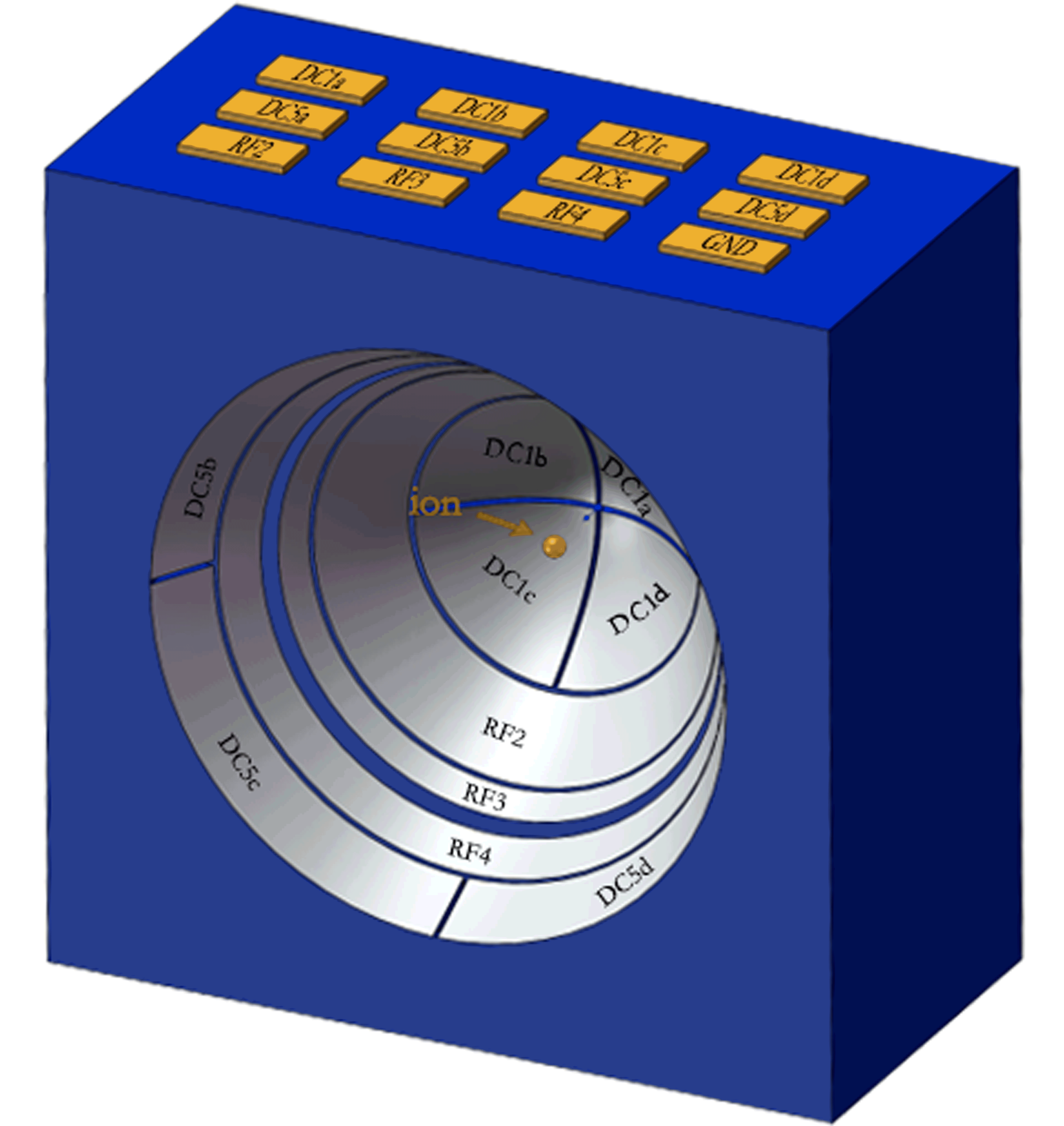}
	 \caption{\label{bulktrappad} The quantum network node package design scheme within the parabolic-mirror electrodes. The outer blue cladding is made of insulating material (e.g. silica) to support the main trap system. The inner gray paraboloid is aluminum coated and engraved used as electrodes. Twelve different electrode pads are set on the top of the model in brown substrates. They correspond to the labels inside respectively from RF2-4, DC1a-1d and DC5a-5d. An orange ion for demonstration is put on the $z$ axis focus of our system. Note that for clarity, we adjust the proportion of different parts of our model.}
 \end{figure}

 \begin{figure*}[htbp]
	\centering 
	\subfigure[]{
		\includegraphics[width=8.6cm]{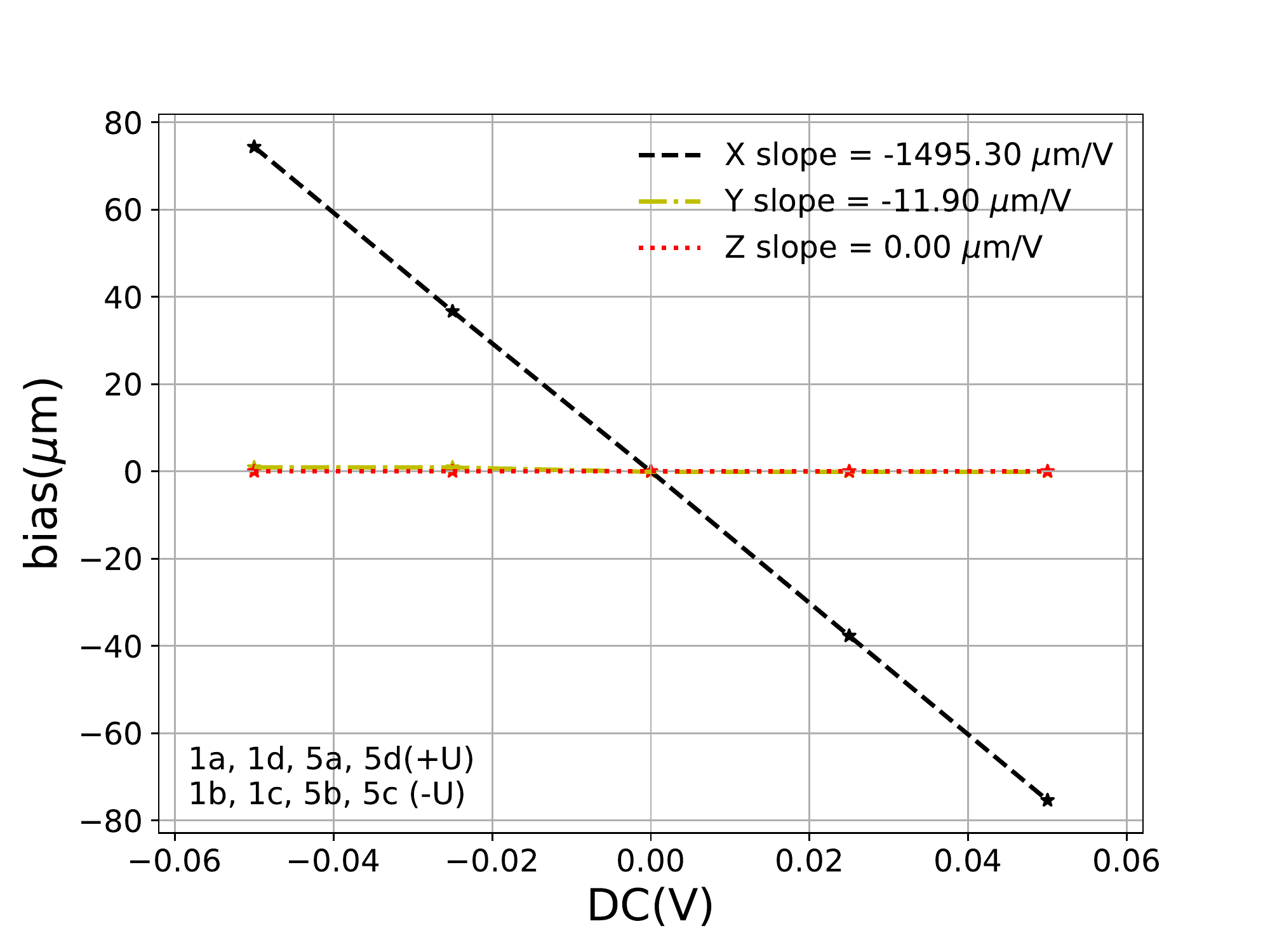}
	}%
	\subfigure[]{
		\includegraphics[width=8.6cm]{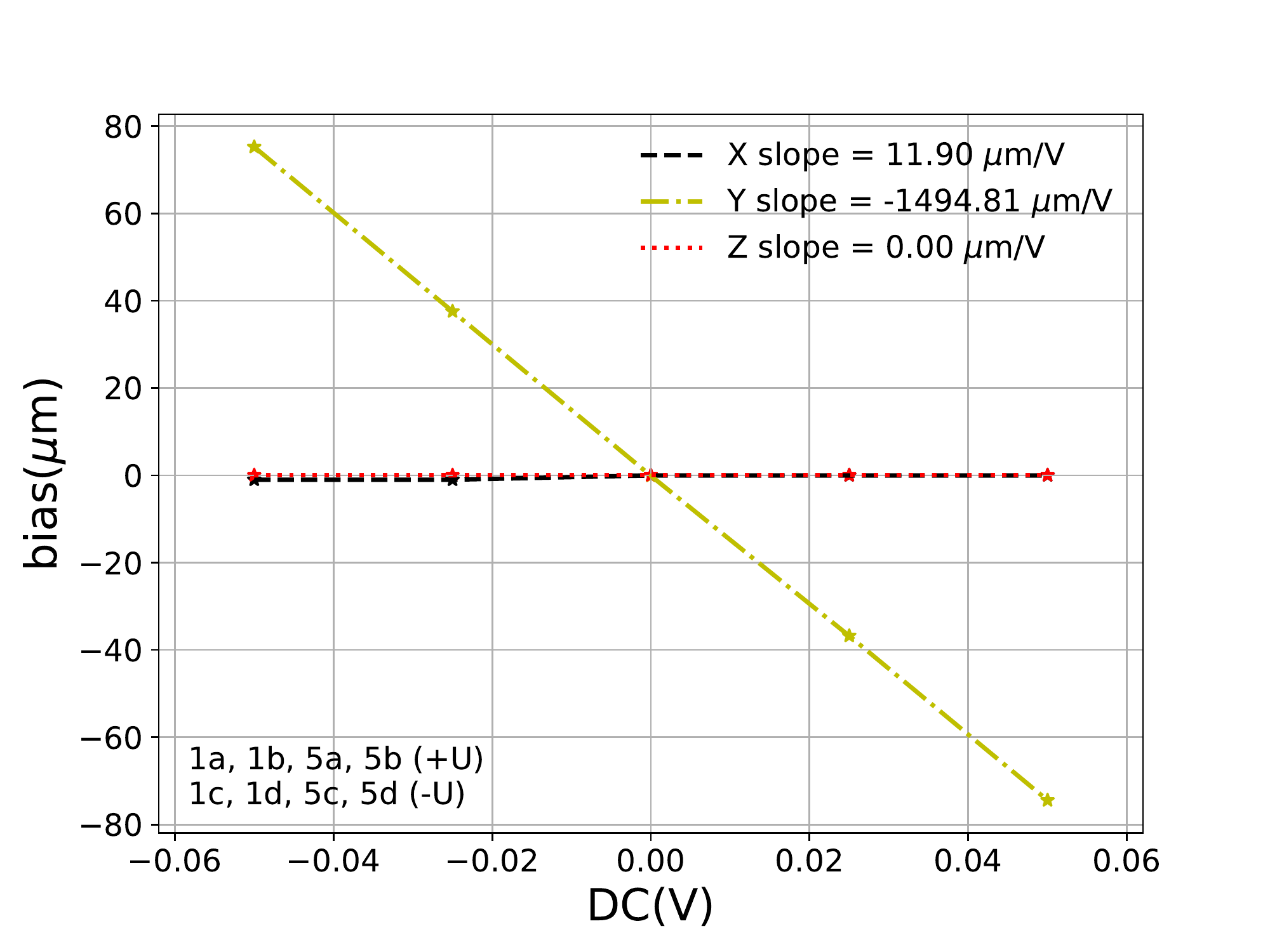}
	}
	\subfigure[]{
		\includegraphics[width=8.6cm]{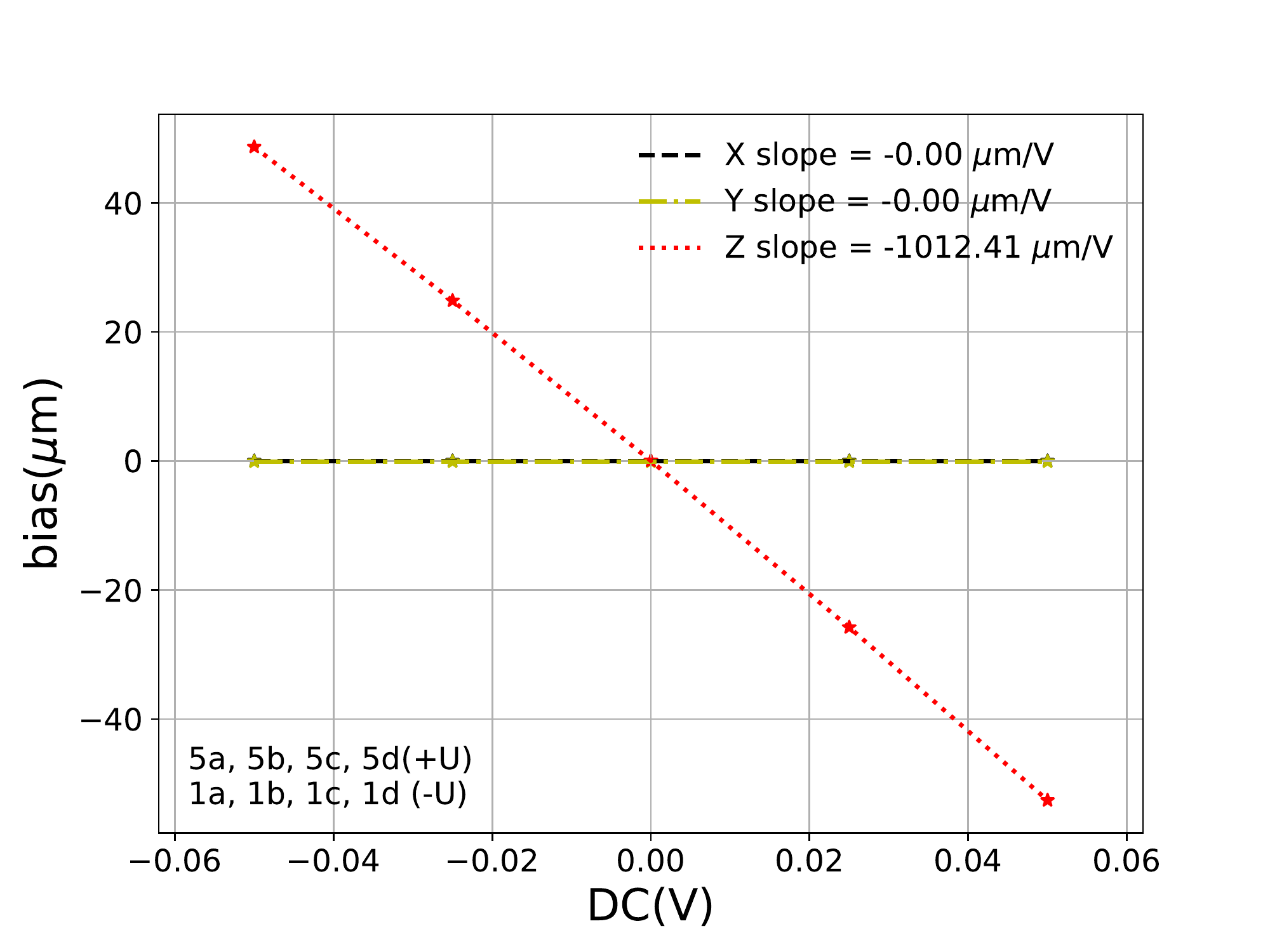}
	}\quad
	\subfigure[]{
		\includegraphics[width=8.3cm]{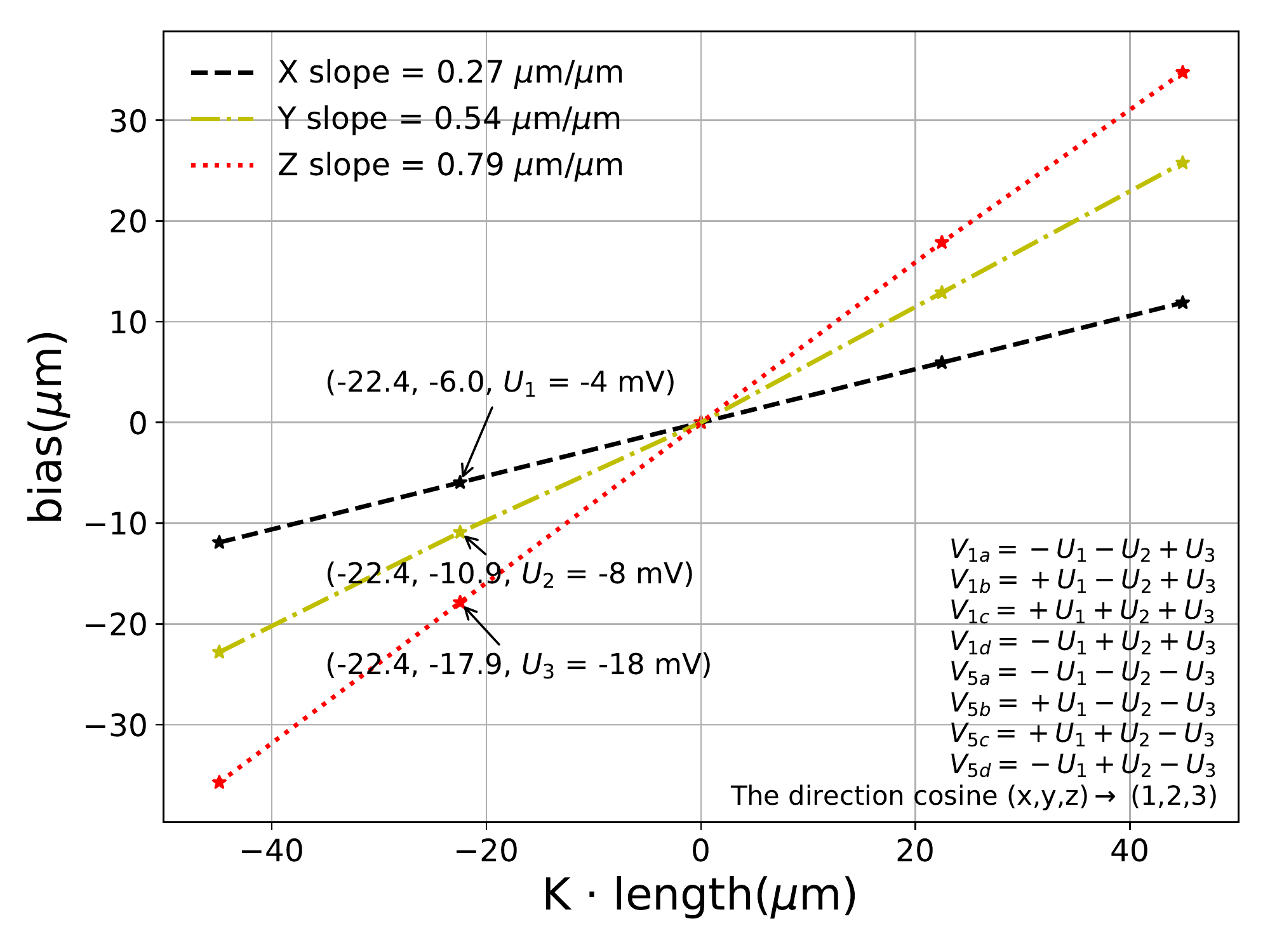}
		\label{fig:3d} 
	}%
	\caption{\label{DC_individual} The DC compensating voltage strategy. A pair of compensating voltage $\pm U_i \ (i=1, 2, 3)$ is applied on two opposite electrodes. Each compensating direction is linear and independent. (a-c) When electrodes a and d (including 1a, 5a, 1d, and 5d) are applied $+U_1$ voltage at the same time, and electrodes b, c (including 1b, 1c, 5b, and 5c) applied $-U_1$ voltage, the ions move independently in the $-x$ direction. The moving distance in the $y$ and $z$ directions is zero, and the relationship between the moving distance and the voltage is $-1495 \,\mathrm{\mu m/V}$; In a similar way, the applied voltage $\pm U_2$ and $\pm U_3$ will individually move ion along $-y$ or $-z$ direction, the ratio is $-1495 \,\mathrm{\mu m/V}$ or $-1052 \,\mathrm{\mu m/V}$ respectively. (d) shows that the DC compensation strategy is linear and independence. One of the three testing voltages $U_1=0.669\,\mathrm{mV}, U_2=1.338\,\mathrm{mV}, U_3=2.852\,\mathrm{mV}$ can be calculated with the assumed movement along the direction $x=1\um$, $y=2\um$, and $z=3\um$. The simulated ion displacement direction and the assumed moving distance are almost the same.}
\end{figure*}

The trap structure of this design will employ a metal reflection surface as the electric electrodes~\cite{luo2009protocols}, using the coating of structured aluminum metal film on silica based material. The gold electric pads are on the top side, shown in Fig.~\ref{bulktrappad}. The dimension of the parabolic shape electrodes is shown in Fig.~\ref{frontview} and Fig.~\ref{topview}. For convenience, we define the vertex of the paraboloid of the revolution as the origin of coordinates, the paraboloid axis of rotation symmetry as the $z$-axis, and the two other orthogonal directions as $x$-axis and $y$-axis, respectively. The equation of the paraboloid is $z(r)=r^2/4f$, where $r\equiv \sqrt{x^2+y^2}$ is the radial coordinate in the $x$-$y$ plane, and $f=2.1\mm$ is the focal length. The depth and front opening diameter of the paraboloid are $29.5\,\mathrm{mm}$ and $31.5\,\mathrm{mm}$, respectively. The paraboloid is divided into 5 independent segments, segments 1 and 5 are DC electrodes, and segments 2-4 are RF trap electrodes. The dimensions of the RF and DC electrodes are shown in Fig.~\ref{topview} respectively. When the parameters in Table~\ref{basis_para} are applied, the position of the extremum of the potential on the $z$ axis exactly coincides with the saddle point. The shift of the RF saddle point along the axis in $z$ direction can be realized by changing the voltages on three RF electrodes of V2-V4. In order to compensate the ion micromotion and adjust the equilibrium position when the ion is stably trapped, the two DC electrodes of electrode 1 and electrode 5 can be further divided into 4 independent parts, respectively labeled as 1a-1d and 5a-5d. As shown in Fig.~\ref{topview}, electrodes a-d are located in quadrants I-IV of the coordinate system, respectively. The relationship between the distance that the ion moves and the change of voltage is shown in Fig.~\ref{DC_individual}(a-c). 

We adopt a method of applying three voltage pairs ($\pm U_{1}$, $\pm U_{2}$, $\pm U_{3}$) to three group of electrodes to control the ion equilibrium along three individual directions at the same time. Then the actual voltage applied to one DC compensation electrode is the sum of the three pairs of voltages on that electrode. The relationship between actual voltage applied on electrode and the group voltage is shown in figure~\ref{fig:3d}. 
In order to verify that the movements in the three directions are independent and linearly additive, we test whether the actual movement direction of the ion is the same as the supposed direction vector and the movement distance $l=(m_1,m_2,m_3)$. The simulated results prove that the above DC compensation scheme works very well and is independent in the three directions. This is demonstrated in Fig.~\ref{fig:3d}, where the horizontal axis stands for the assumed ion move distance along the direction-vector $(1,2,3)$, and the vertical axis stands for the distance that the simulated ions move in the each directions. The slope ratio of the three directions is nearly the same as 1:2:3, satisfying the independent relationship.

\begin{figure*}[htbp]
 	\centering 
 	\subfigure[]{
 		\includegraphics[height=6.5cm]{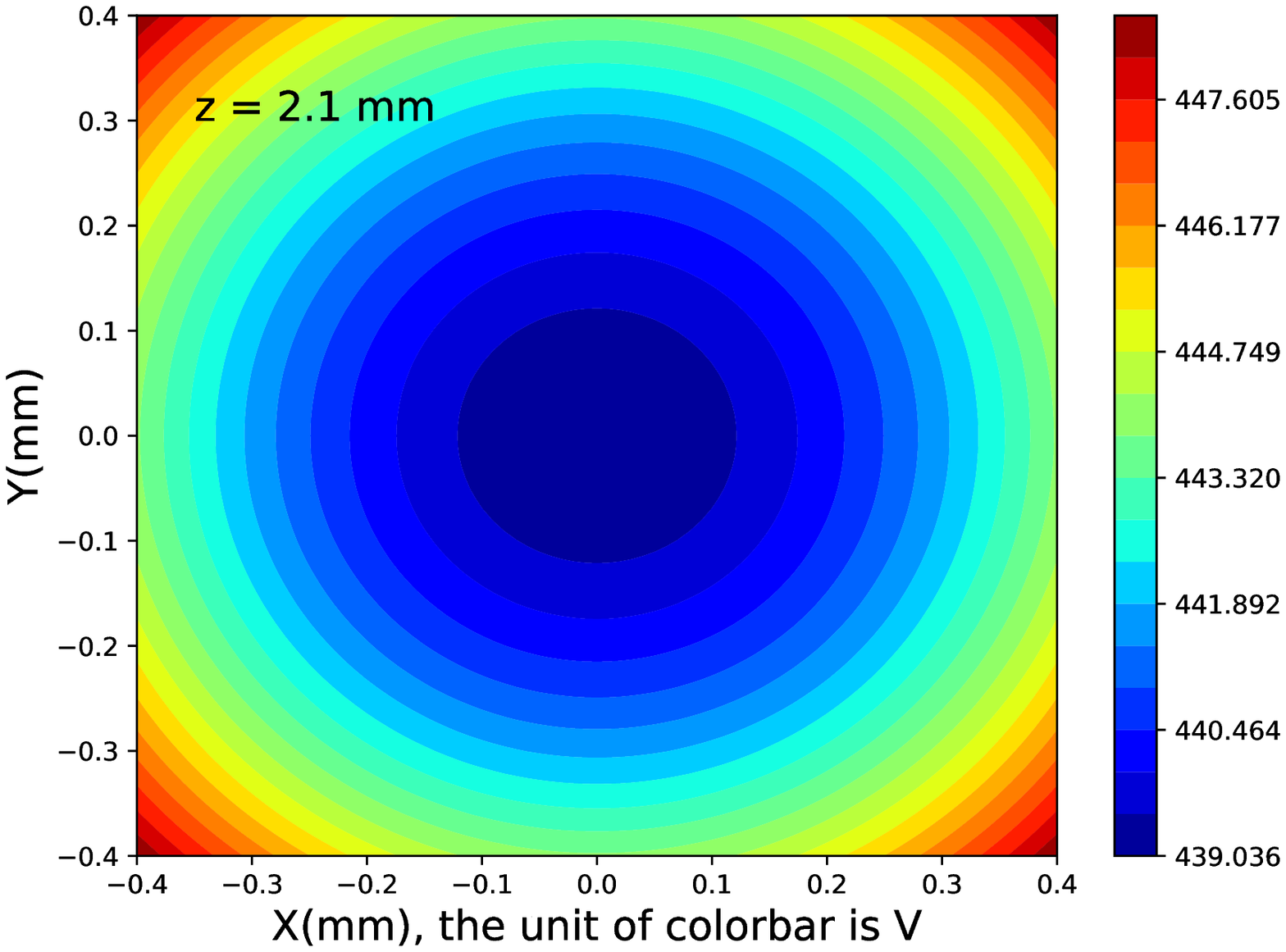}
 		\label{fig:4a} 
 	}%
 	\subfigure[]{
 		\includegraphics[height=6.cm]{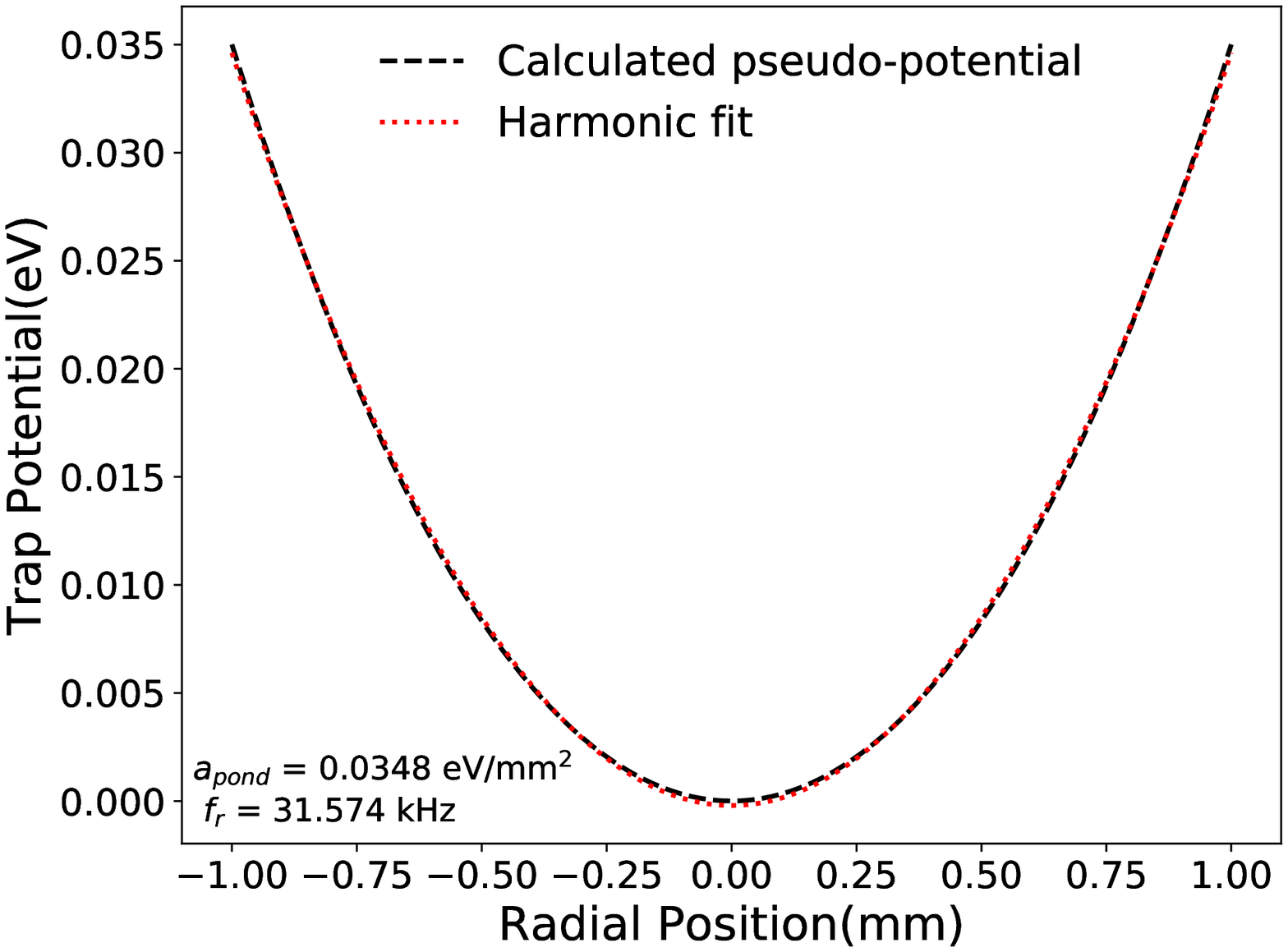}
 	}
 	\subfigure[]{
 		\includegraphics[height=6.5cm]{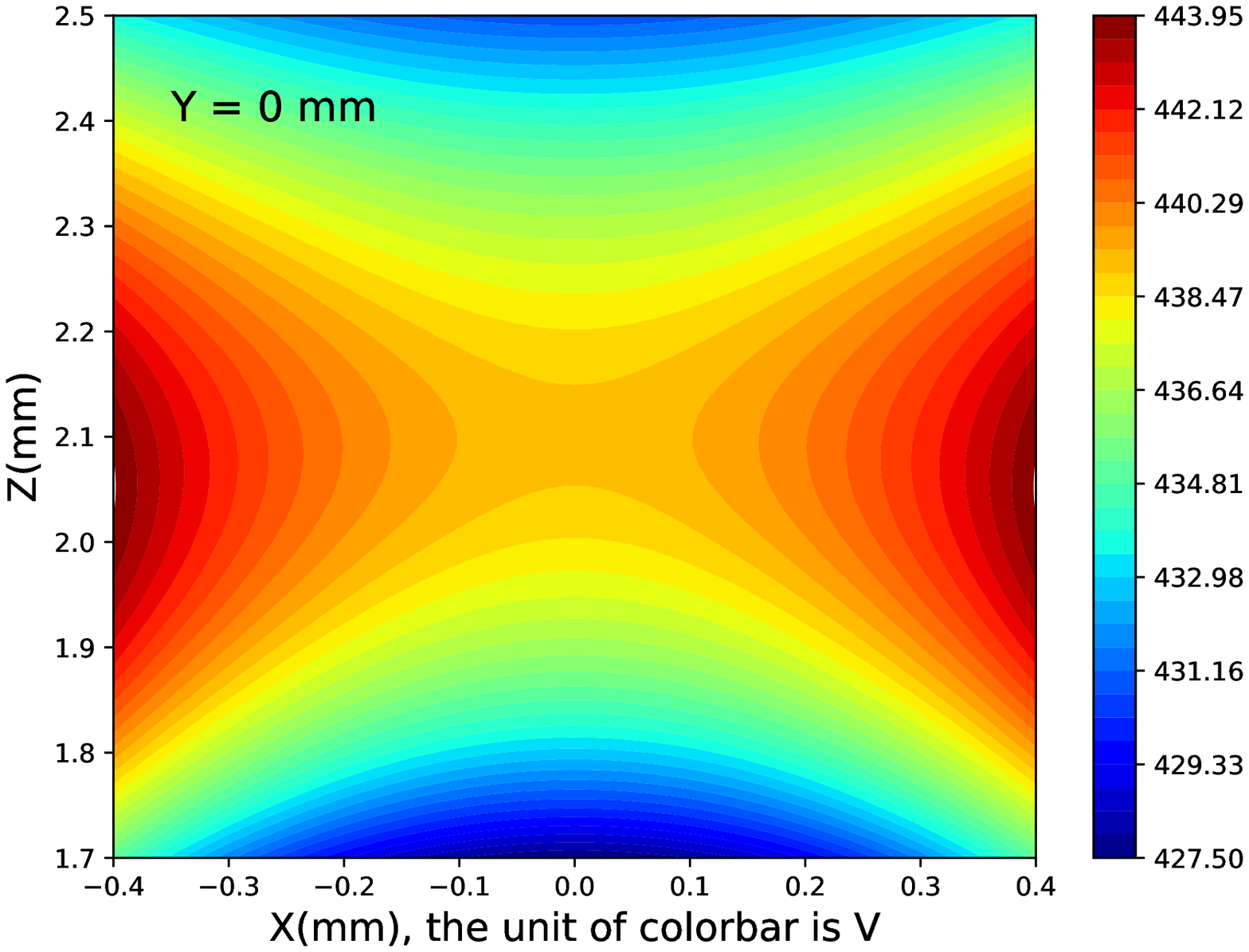}
 		\label{fig:4c} 
 	}%
 	\subfigure[]{
 		\includegraphics[height=6.cm]{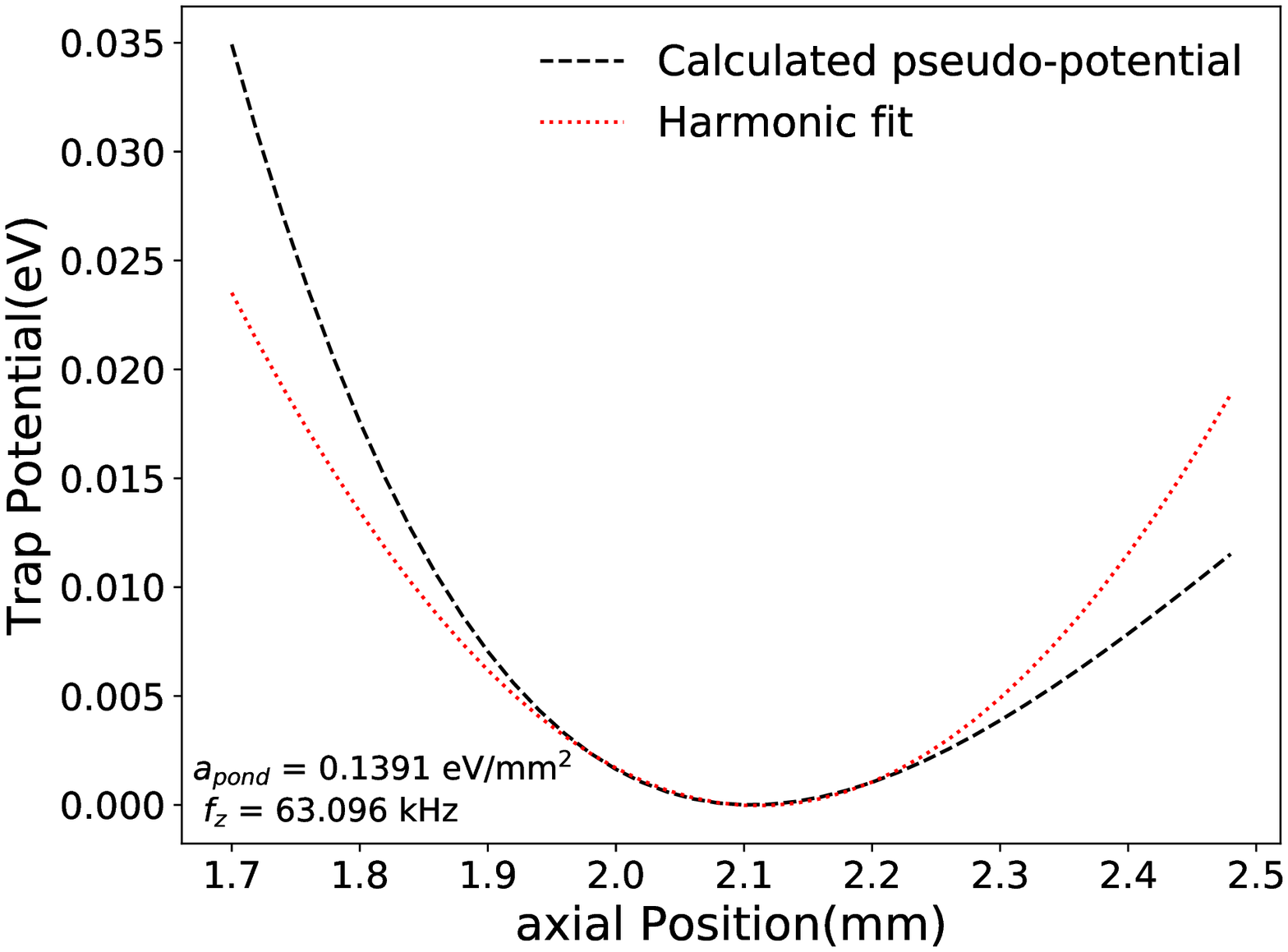}
 	}%
 	\caption{\label{pso} The simulation results of the electric field and the calculated trap parameters. (a)(c) The electric potential in the XoY plane and XoZ plane. The electric field is distributed as an axial-symmetric paraboloid in XoY plane and the electric potential isosurface is a saddle shape. (b)(d) the pseudo-potential is fitted to $0.035\,\mathrm{eV}$ while $r \in (-400,400) \um$, and the secular motion frequency is $31.574 \kHz$. The pseudo-potential $0.003\,\mathrm{eV}$ along z direction is fitted in the range of $(1950,2250) \um$, and the secular motion frequency is $63.096 \kHz$.}
\end{figure*}

Using electrostatic/magnetic fields numerical simulation software CPO (Charged Particle Optics)\cite{read2011cpo,read2015achieving}, we calculated the electric field potential distribution of the parabolic-mirror trap in the (XoY) plane (Fig.~\ref{fig:4a}) and the (XoZ) plane (Fig.~\ref{fig:4c}). From the potential results, we further calculated pseudo-potential and secular motion frequency in radial ($x$ and $y$) and axial ($z$) directions according to Eq.~\eqref{w_D} ~\cite{marquez2016novel}. 
In Eq.~\eqref{w_D}, $\bar{D}_\text{pseudo}$ is the calculated pseudo-potential, $Q$ is ion's charge, $V_0$ is the distribution of real potential, m is the mass of the ion, $r_0$ is the minimum distance between the ion and RF electrodes, $\Omega$ is the RF driven frequency, $\omega$ is the secular motion frequency of the ion. 
In the radial direction, the quadratic fitting coefficient $a_{\rm pond}$ is 0.0348 eV/mm, and the secular frequency $f_r = \omega_r/2\pi = 31.6 \kHz$. 
In the axial direction, the quadratic fitting coefficient $a_{\rm pond}$ is 0.1391 eV/mm, and the secular frequency $f_z = \omega_z/2\pi = 63.1 \kHz$.

\begin{equation}
	\bar{D}_\text{pseudo}=\frac{Q V_{0}^{2}}{4 m r_{0}^{2} \Omega^{2}}
	,\quad
	\omega=\sqrt{\frac{2 q a_\text{pond}}{m}}
	\label{w_D}
\end{equation}

\begin{table}[htbp] 
	\caption{the parameters of the parabolic mirror trap and $^{171}\Yb^+$ ion.}  
	\begin{tabular}{c p{3cm}<{\centering} c} 
		\toprule[1pt] 
		Notation & Value & Comment\\
		\midrule[0.5pt]
		\text{$V_1$}  & 0.35\,V   &\\  
		\text{$V_2$}  & 819.20\,V &\\ 
		\text{$V_3$}  & 541.00\,V &\\  
		\text{$V_4$}  & 712.75\,V &\\ 
		\text{$V_5$}  & 0.50\,V \\ 
		$m$       & 171\,u  & ion's mass \\ 
		$Q$     & 1\,e      & ion's charge \\  
		$\Omega$   & 20\,MHz & RF frequency\\ 
		\bottomrule[1pt]
	\end{tabular}  
	\label{basis_para}
\end{table}

\section{Optical Properties \label{sec:prop}}

\begin{figure}[htbp]
	\includegraphics[width=.5\textwidth]{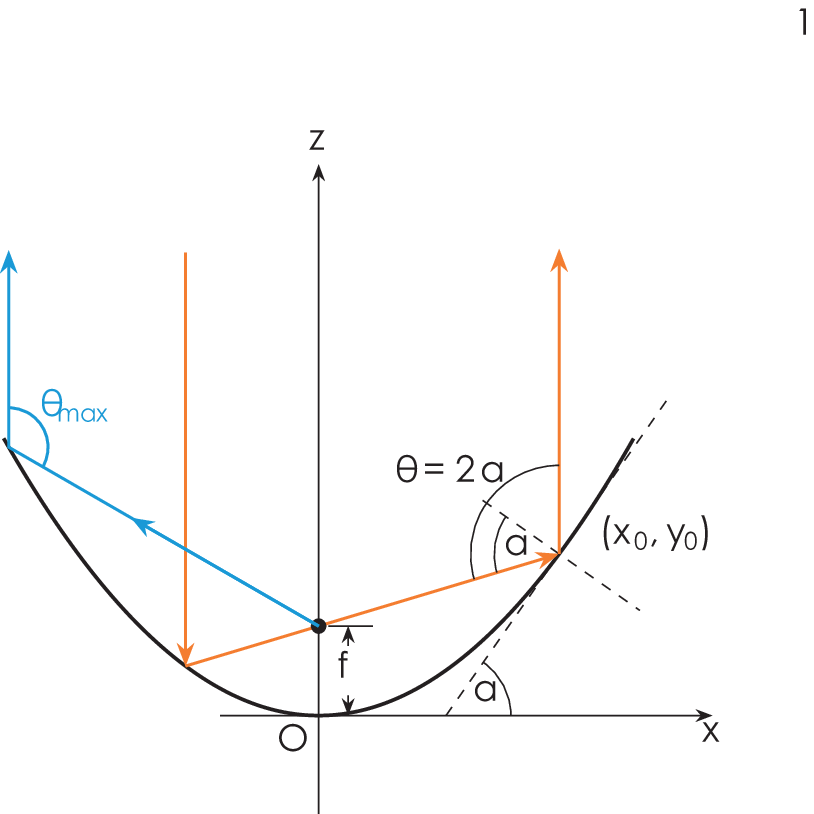}
	\caption{\label{optischeme} The optical scheme of the paraboloid surface for photon collection. The ion at the focus is emitting photons. The photons are reflected vertically by the parabolic surface. The photon incident angle is $\alpha$, and the light deflection angle is $\theta=2\alpha$. The maximum reflection angle $\theta_{\mathrm{max}}$ is decided by the parabolic-mirror aperture. In this paper design, the $\theta_{\mathrm{max}}=150^\circ$.}
\end{figure}

\begin{figure}[htbp]
	\includegraphics[width=.47\textwidth]{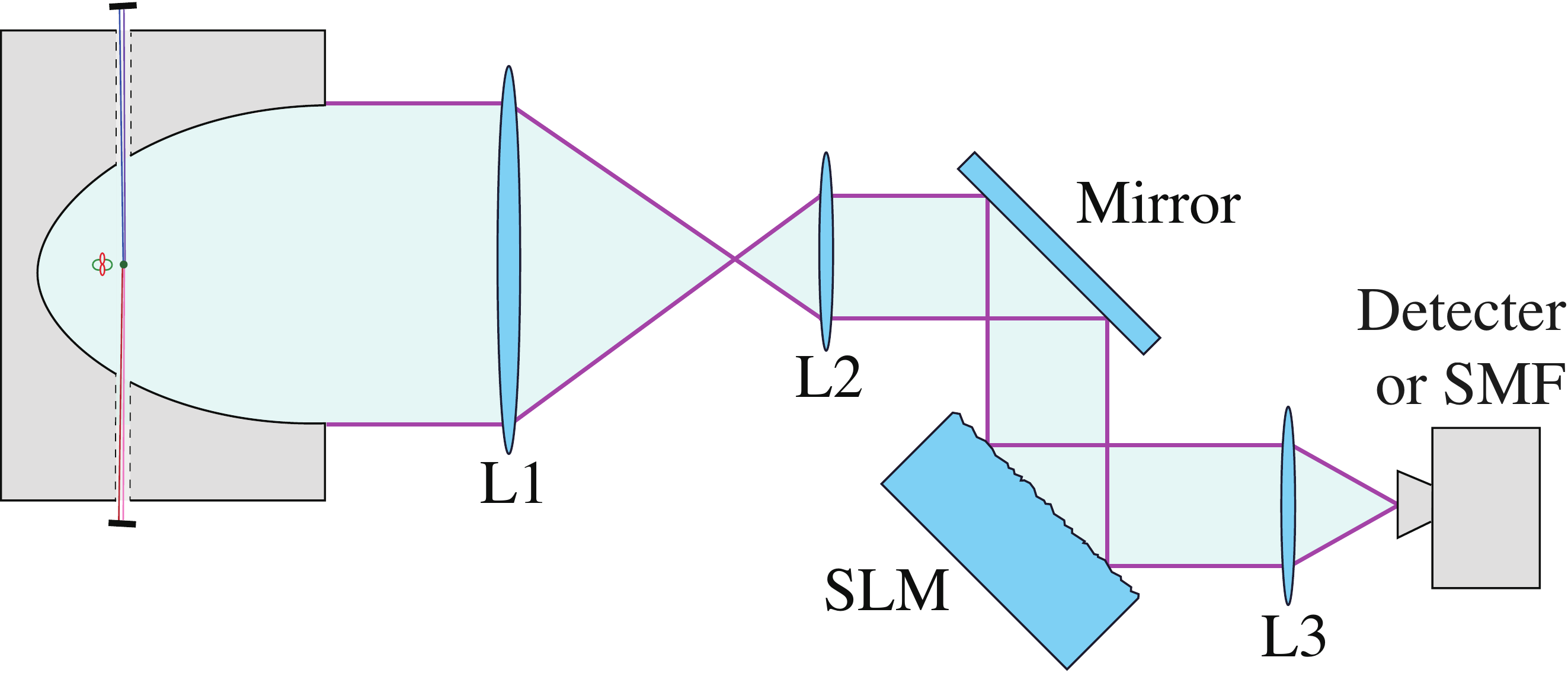}
	\caption{\label{optpath}The laser optics scheme. The ion is an dipole-emission light source, 
	trapped at the focus of the parabolic-mirror. The cooling and operational laser ($369\,\mathrm{nm}$/$399\,\mathrm{nm}$/$638\,\mathrm{nm}$/$935\,\mathrm{nm}$) exciting the ion from the side of the paraboloid surface, perpendicular with the $z$ axis. The photon collection, beam shaping and detection functions are shown in the optical path. With wavefront modulator device(e.g. SLM), the emitting dipole pattern can be transformed to Gaussian mode, and coupled into single mode fiber with a high efficiency}
\end{figure}

The scattering fluorescence from the ions trapped at the focal point is reflected and collected by the paraboloid. The scheme is shown in Fig.~\ref{optischeme}, the light from the focal point bounces off the paraboloid and becomes parallel light that shoots out.
Let the coordinate of the reflection point where the ray from the focus intersects the paraboloid be $(x_0,y_0)$, and the slope of the tangent line is $\alpha$ ($\alpha$ is also the angle at which light reflects off the paraboloid).
Thus, the light is reflected by the paraboloid and then emitted along the $+z$ axis. The deflection angle of the light is $\theta=2\alpha$.
The maximum incidence angle that can be reflected by the paraboloid depends on the size of the opening of the paraboloid. The angle of deflection of the light reflected by the opening edge of the paraboloid is $\theta=150^\circ$. Therefore, the paraboloid can not collect photons only within the range of $75^\circ<\alpha<90^\circ$, and the corresponding space collecting solid angle is ${(1-\cos\theta)}/{2}=93.3\%$.

In the experiment of $^{171}\Yb^+$ ion trapping, multiple lasers are used ($369\,\mathrm{nm}$ cooling/pumping/detecting lasers, $399\,\mathrm{nm}$ ionization laser, $638\,\mathrm{nm}$ and $935\,\mathrm{nm}$ repump lasers), and the atomic beam is forced into the trap, both are from the gap between the 3rd and 4th electrodes, and the width of the gap is 200$\um $. The focus is in the middle of the gap between the 3rd and 4th electrodes, and the distance between the focus and the electrode gap is $2.1\mm$. The $399\,\mathrm{nm}$ and $369\,\mathrm{nm}$ laser will be aligned into the trap in the XoY plane $z=f=2.1\mm$ in the direction perpendicular to the direction of atomic beam, and the $638\,\mathrm{nm}$ and $935\,\mathrm{nm}$ repump lasers can also be aligned into the parabolic well in the plane, Fig.~\ref{optpath}.

The laser beam parameters can be calculated from the electrode $256\um$ gap width and the $2.1\mm$ distance between electrodes and ions: if the $369\,\mathrm{nm}$ laser at the ions is focused to a $50\um$ beam waist, then Rayleigh length of $21.28\mm$ can be used. The beam diameter of the laser at the electrode gap is $100.5\um$, which is less than the gap width of $256 \um$. Therefore, the state initialization, cooling, detection and other operations required by $\Yb^+$ ions can be achieved.
During state detection, the fluorescence is reflected by the paraboloid and then propagates as parallel light, and then focused by a lens onto the CCD for observation, or coupled to a single-mode fiber, as shown in Fig.~\ref{optpath}. For individual addressing, the parabolic mirror is used reversely, shooting laser back from the collection direction with small angle tilting.

In the quantum network scheme for ion-ion entanglement, it is necessary to select the mode of the collected photons. This can be achieved by coupling the collected photons into a single-mode fiber, and the coupling efficiency is also an important factor in determining the overall optical collection rate of the system~\cite{kim2011efficient}. Ref.~\cite{sondermann2007design} indicates that the dipole radiation field after the parabolic-mirror reflection corresponds to the orbital angular momentum(OAM) $\LG_{10}$ mode in space. Using the OAM light theory, $\LG_{10}$ mode can be transformed into $\LG_{00}$ Gaussian mode through a Spacial Light Modulator(SLM) or a phase plate. The Gaussian mode is also the intrinsic mode of single-mode fiber, which can theoretically realize a perfect coupling efficiency\cite{kim2011efficient}. On the other hand, wavefront modulating device, like SLM, can correct the aberration in the optical path besides the function of mode conversion~\cite{golla2012generation,chou2017note}. Finally, the total light collection efficiency of the system equals $P_\Omega \cdot R \cdot P_{f} \cdot P_t= 93.3\% \cdot 90\% \cdot 90\% \cdot 95\%=71.8\%$ (where ${P_\Omega}$ represents the solid angle, $R$ represents the reflection coefficient, $P_{f}$ represents the fiber coupling efficiency, and $P_t$ represents the transmission efficiency), which is 7.23 times higher than the situation of 10\% collection efficiency. In the two-photon ion entanglement scheme, the probability of success is proportional to the square of the light collection efficiency~\cite{Duan:2004:STI:2011617.2011618}, and it can be estimated that system entanglement can reach $183\mathrm{Hz}\cdot 7.55^2= 10.26\kHz$, if this design is applied on the experiment~\cite{stephenson2019high}. Regarding the reflectivity of the parabolic mirror, it can be seen from the query that the reflectance of the UV-enhanced aluminum coating to the 369nm wavelength can exceed $90\%$.

\section{Spatial Coincidence of Focus and RF Null Point\label{sec:overlap}}

 \begin{figure}[htbp]
 	\includegraphics[width=9cm]{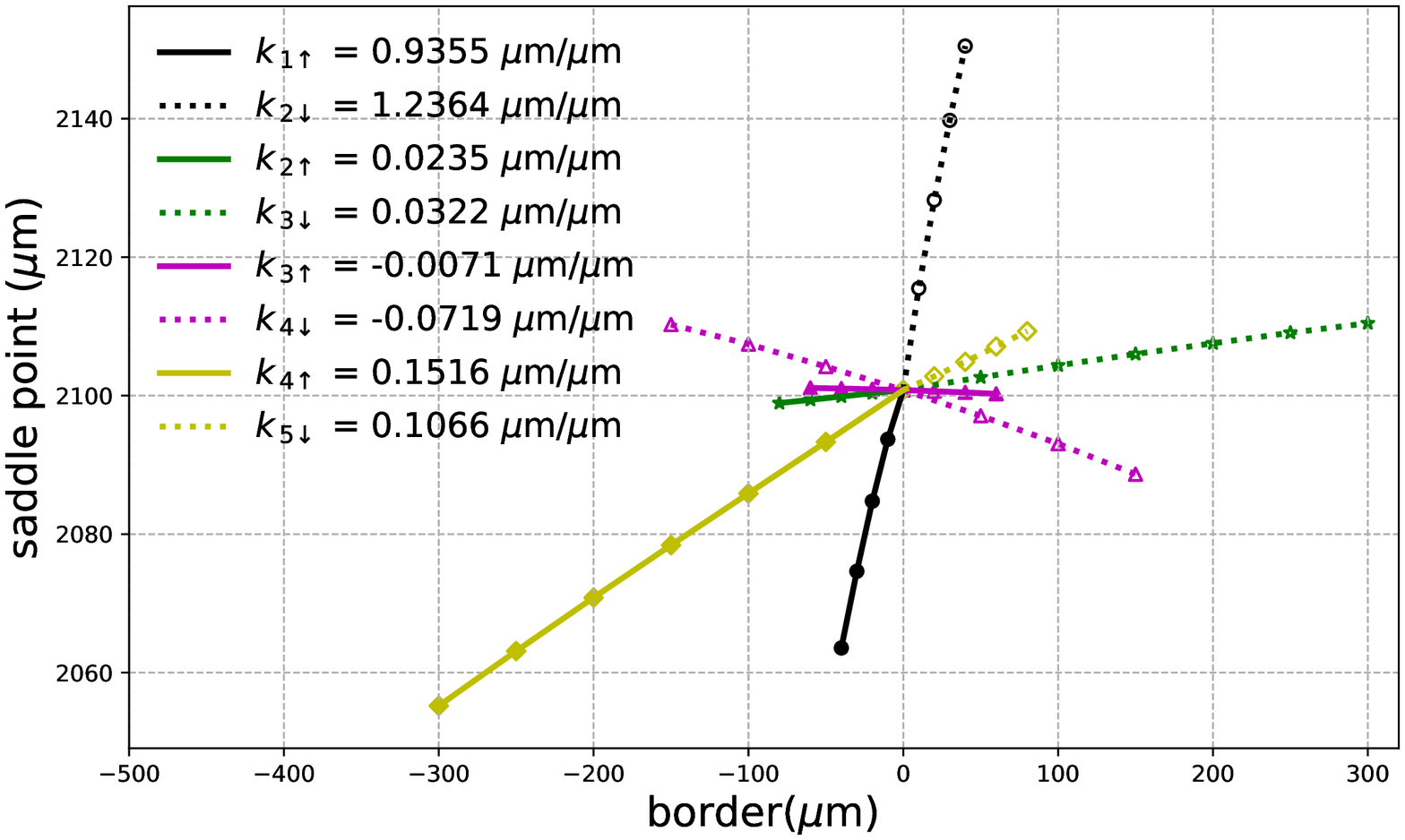}
 	\caption{\label{nullconstr} The sensitivity analysis was shown here between the structure dimension and RF null point. The label describes the electrode number, and the perturbing edge, e.g. the $3^\text{rd}_\downarrow$ means the 3rd electrode segment's bottom edge (the edge at $z=1.136 \mm$), etc. The RF null point shift ratios are: $0.9355\um/\um$ for the $1^\text   {st}_\uparrow$, $1.2364\um/\um$ for the $2^\text{nd}_\downarrow$, $0.0235\um/\um$ for the $2^\text{nd}_\uparrow$, $0.0322\um/\um$ for the $3^\text{rd}_\downarrow$, $-0.0071\um/\um$ for the $3^\text{rd}_\uparrow$, $-0.0719\um/\um$ for the $4^\text{th}_\downarrow$, $0.1516\um/\um$ for the $4^\text{th}_\uparrow$, $0.1066\um/\um$ for the $5^\text{th}_\downarrow$. }
 \end{figure}

 \begin{figure}[htbp]
 	\includegraphics[width=9cm]{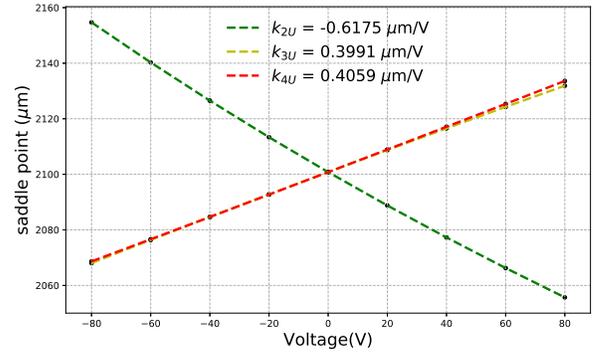}
 	\caption{\label{nullvolt} The sensitivity analysis was shown for the relationship between voltage and RF null point. The setting RF voltage is given followed by a voltage label in the bracket, e.g. $V_{2}(\tilde{V}_2=819V)$, etc. The results are: $-0.6175\,\mathrm{\um/V}$ for the the $V_{2}(\tilde{V}_2=819V)$ segment, $0.3991\um/V$ for the the $V_{3}(\tilde{V}_3=541V)$ segment, $0.4059\um/V$ for the the $V_{4}(\tilde{V}_4=708V)$ segment.}
 \end{figure}

Using software CPO, we simulate and analyze the sensitivity of the trap RF saddle point position by perturbing the structure dimension and the RF voltage. The results are shown in the Fig.~\ref{nullconstr} and Fig.~\ref{nullvolt}. The RF null position is recorded as setting different RF electrode segment border size each. The $x$ axis is the border size variation (in unit of $\um$). A positive (negative) border variation means that the segment border shifts in the $+z$ ($-z$) direction.  For example, as the 3rd electrode border size varies $+1\um$, the RF saddle point shifts $0.0322\um$ along the $+z$ axis. The dependence of the RF saddle point position on the RF voltage variation is shown in Fig.~\ref{nullvolt}, where the horizontal axis stands for the RF voltage difference from the setting RF voltage, and the vertical axis stands for the saddle point position shift. If the output voltage of RF signal can be controlled within $2\,\mathrm{V}$, the RF saddle point can be shifted within $1\um$ precision. 

The relations between the RF saddle point shift and electrode segment deformations are linear and independent, and so is the relation between RF saddle point shift and the RF voltage variation. The final RF saddle point position can be described by the following fitting formula 
\begin{equation}
y_{\text{saddle}}=y_{0}+\d y_{\text{vt}}+\d y_{\text{st}},
\label{longeq1}
\end{equation} 
where $y_{\text{saddle}}$ is the actual position of the saddle point with all of the electrodes' effects; $y_0$ is the ideal position of the saddle point, here the focal point is at $y_0=2.1 \mm$; $\d y_\text{vt}$ is the saddle point displacement due to voltage variation; and $\d y_{\text{st}}$ is the saddle point displacement due to electrodes size variation.Eq.~\eqref{longeq2} gives a linear approximation to estimate the relation between $\d y_\text{vt}$($\d y_{\text{st}}$) and electrodes' voltage (dimension) variation.

In Eq.~\eqref{longeq2}, $k_{iU}$ is the linear slope of the displacement of saddle point by changing the $i$-th segment's voltage (here and hereafter $i\in\{2,3,4\}$); Similarly, the parameter $k_{i\downarrow}$ corresponds to the dependence on the lower edge dimension of the $i$-th segment structure, and $k_{i\uparrow}$ is the upper edge dimension of the $i$-th electrode. The symbols ${V}_{i}$, $x_{i\downarrow}$ and $x_{i\uparrow}$ represent the $i$-th segment's voltage, the $i$-th segment's coordinate of the lower and the upper edge, respectively. 
And the symbols $\tilde{V}_{i}$, $\tilde{x}_{i\downarrow}$ and $\tilde{x}_{i\uparrow}$ represent the reference parameters when the $i$-th segment of the voltage, lower and upper edge have the specified values in Appendix~\ref{ap:param}, as well as the parameters used in the linear relationships. 



To compare the displacement effects of the voltage and the electrodes size variation, the fitting slope parameters $k_{i\uparrow(\downarrow)}$ and the $k_{i {U}}$ are in the same level. Thus, the RF null point displacement induced by voltage and the structure variations are in the same level, i.e. we can always compensate or cancel the mismatch between RF null point and the focal point by changing the RF voltage or electrodes size. After the device was fabricated, the electrodes structure are settled. The RF null point can still be shifted and aligned with the focal point by accurately adjusting the three RF voltages. The other advantange of using three RF driven voltages is that the machining accuracy of the trap structure can be relaxed by adjusting electrodes' voltages.

\section{Summary\label{sec:sum}}
In summary, we present a design of a parabolic-mirror ion trap, which has a very high photon collecting solid angle and the ability to make the parabolic focus and the RF null position precisely coincide, by changing three electrodes' RF voltage. This design is manufacture-friendly because it requires less machining accuracy. By integrating the electric electrodes and the photon collection device, the design is compact and easy to fabricate. Moreover, the design has a good potential to greatly increase the collecting photon efficiency after SMF filtering and improve the imaging quality to the diffraction limit. Finally, we expect the ion-ion entanglement generation speed can be improved dramatically by using this design. 

\section*{Acknowledgements}
This work was supported by the Fund of Natural Science Foundation of Guangdong Province of China (Grants No. 2017A030310452 and 2020A1515010864), Guangdong Basic and Applied Basic Research Foundation (Grants No. 2020A1515010864) and the National Natural Science Foundation of China (Grants No.11904423 and 11774436), the Key-Area Research and Development Program of Guang- Dong Province under Grant No.2019B030330001, Guangdong Province Youth Talent Program (Grant No.2017GC010656), National Training Programs of Innovation and Entrepreneurship for Undergraduates(Grants No.201901139). Sun Yat-sen University Foundation for Youth (Grants No.17lgpy27). Authors thanks Feng Zhu for discussion.

\subsection*{Author information}
These authors contributed equally: Ben-Ran Wang, Qing-Lin Ma, Jia-Yu Guo, Ming-Shen Li, Yu Wang

\appendix{}
\section{Linear Parameters}\label{ap:param}

The symbols $\tilde{V}_{i}$, $\tilde{x}_{i\downarrow}$ and $\tilde{x}_{i\uparrow}$ represent the reference parameters when the $i$-th segment of the voltage, lower and upper edge have the specified values in Eq.~\eqref{longeq3}. 
And all of the parameters used in the linear relationships are listed in Eq.~\eqref{longeq3}. The range of parameters are shown in the bracket form.

\begin{equation} 
	\left\{
	\begin{array}{ll}
	\d y_{\text{vt}}=
	& k_{2 {U}}(V_{2}-\tilde{V}_{2 })+k_{3 {U}}(V_{3 }-\tilde{V}_{3  }) \\
	& +k_{4{U}}(V_{4}-\tilde{V}_{4} ) \\
	\d y_{\text{st}}=
	& k_{2 \downarrow}\left(x_{2  \downarrow}-\tilde{x}_{2\downarrow}\right)+k_{2 \uparrow}\left(x_{2 \uparrow}-\tilde{x}_{2 \uparrow}\right)\\
	& +k_{3 \downarrow}\left(x_{3  \downarrow}-\tilde{x}_{3\downarrow}\right)+k_{3 \uparrow}\left(x_{3 \uparrow}-\tilde{x}_{3 \uparrow}\right)\\
	& +k_{4 \downarrow}\left(x_{4\downarrow}-\tilde{x}_{4\downarrow}\right)+k_{4\uparrow}\left(x_{4\uparrow}-\tilde{x}_{4\uparrow}\right)\\
	& +k_{5 \downarrow}\left(x_{5 \downarrow}-\tilde{x}_{5\downarrow}\right)+k_{1 \uparrow}\left(x_{1 \uparrow}-\tilde{x}_{1 \uparrow}\right)
	\end{array}
	\right.
	\label{longeq2}
	\end{equation}
	
	\begin{equation} \label{longeq3}
	\left\{	
	\begin{array}{ll}
	y_0=2.100\mm, &\tilde{x}_{1\uparrow}=0.256\mm,\\
	\tilde{V}_{2}=819.20\V, &\tilde{x}_{2\downarrow}=0.296\mm, \\
	\tilde{V}_{3}=541.00 \V, &\tilde{x}_{2\uparrow}=1.096\mm, \\
	\tilde{V}_{4}=708.00\V, &\tilde{x}_{5\downarrow}=3.040\mm, \\
	\tilde{x}_{3\downarrow}=1.136\mm, &\tilde{x}_{3\uparrow}=1.900\mm, \\
	\tilde{x}_{4\downarrow}=2.300\mm, &\tilde{x}_{4\uparrow}=3.000\mm, \\
	k_{2U}=-0.6175 \um/\V, &V_2\in(739.2-899.2)\V, \\
	k_{3U}=0.3991  \um/\V, &V_3\in(461.0-621.0)\V, \\
	k_{4U}=0.4059  \um/\V, &V_4\in(628.0-788.0)\V, \\
	
	k_{1\uparrow}= 0.9355\um/\mu\mathrm{m}, &x_{1\uparrow}\in(0.216-0.256)\mm, \\
	k_{2\downarrow}= 1.2364\um/\mu\mathrm{m}, &x_{2\downarrow}\in(0.296-0.336)\mm, \\
	k_{2\uparrow}=0.0235\um/\mu\mathrm{m}, &x_{2\uparrow}\in(1.016-1.096)\mm, \\
	k_{3\downarrow}= 0.0322\um/\mu\mathrm{m}, &x_{3\downarrow}\in(1.136-1.436)\mm, \\
	k_{3\uparrow}=-0.0071\um/\mu\mathrm{m}, &x_{3\uparrow}\in(1.840-1.960)\mm, \\
	k_{4\downarrow}=-0.0719\um/\mu\mathrm{m}, &x_{4\downarrow}\in(2.150-2.450)\mm, \\
	k_{4\uparrow}=0.1516 \um/\mu\mathrm{m}, &x_{4\uparrow}\in(2.700-3.000)\mm, \\
	k_{5\downarrow}= 0.1066\um/\mu\mathrm{m}, &x_{5\downarrow}\in(3.040-3.120)\mm.
	\end{array}
	\right.
	\end{equation}

%

\end{document}